# Positive Narrativity Enhances Sense of Agency toward a VR Avatar

**Kureha Hamagashira[1], Miyuki Azuma[2], Sotaro Shimada[3]**

[1]Graduate School of Science and Technology, Meiji University, Kanagawa, Japan

[2]Organization for the Strategic Coordination of Research and Intellectual Properties, Meiji University, Kanagawa, Japan

[3]School of Science and Technology, Meiji University, Kanagawa, Japan

The full-body illusion (FBI) refers to the experience of perceiving a virtual avatar as one's own body. In virtual reality (VR) environments, inducing the FBI has been shown to modulate users' bodily experiences and behavior. Previous studies have demonstrated that embodying avatars with specific characteristics can influence users' actions, largely through the activation of implicit stereotypes. However, few studies have explicitly manipulated users' impressions of an avatar by introducing narrative context. The present study investigated how avatar narrativity, induced through contextual narratives, affects the FBI. Healthy participants embodied a powerful artificial lifeform avatar in VR after listening to either a positive narrative, in which the avatar used its abilities to protect others, or a negative narrative, in which it misused its power. Participants' impressions of the avatar and indices of bodily self-consciousness were subsequently assessed. The results showed that positive narratives significantly enhanced the sense of agency (SoA), and that SoA was positively correlated with participants' perceived personal familiarity with the avatar. These findings suggest that the avatar narrativity can modulate embodiment in VR.

## 1 Introduction

Individuals operating an avatar in a VR environment may come to experience the avatar's body as if it were their own. This phenomenon is grounded in bodily self-consciousness, the fundamental subjective experience that one's body belongs to oneself. Bodily self-consciousness is commonly conceptualized as comprising two distinct components (Gallagher, 2000). One is the sense of ownership (SoO), which refers to the experience that "this body is mine." For example, viewing one's own back in VR while receiving synchronous tactile stimulation on both the real and virtual body has been shown to induce SoO toward the virtual body (Lenggenhager et al., 2007). The other component is the SoA, defined as the experience that "I am the one who is causing this action." When an avatar's movements are temporally and spatially synchronized with those of the user in a VR environment, users tend to attribute the avatar's actions to themselves, resulting in an enhanced SoA (Kilteni et al., 2012; van Krieken et al., 2017).

Moreover, embodying an avatar with specific characteristics can influence users' own behavior and cognition, a phenomenon known as the Proteus effect (Yee & Bailenson, 2007). For instance, participants embodying a heroic avatar have been shown to exhibit greater prosocial behavior than those embodying a villainous avatar (Yoon & Vargas, 2014). Similarly, embodying an avatar with the appearance of a doctor has been reported to enhance executive function and openness (Koike & Shimada, 2022). Despite these findings, research on the Proteus effect has predominantly focused on influences arising from implicit stereotypes associated with an avatar's appearance, leaving the underlying psychological mechanisms insufficiently understood. Furthermore, individual differences in stereotype endorsement may introduce substantial variability in behavioral outcomes across participants. These limitations suggest that manipulating explicit impressions of an avatar, rather than relying solely on implicit stereotypes, may provide a more controlled and predictable means of modulating users' behavior.

One promising approach for explicitly manipulating impressions of a character is to use media-based narrative approaches to shape how the character is perceived. Characters presented in such media embody narrativity, and this narrativity enables viewers to acquire the character's identity, goals, and perspective—a process referred to as identification (Cohen, 2018). In particular, it has been suggested that when people engage with a narrative, they often internalize a character's thoughts and feelings through the process of identification. In the process of narrative identification, greater liking for a character strengthens identification, and such liking is shaped by audiences' moral evaluations of the character's actions (Cohen, 2018, Raney and Bryant, 2002). Previous studies have shown that identification is promoted when the character is portrayed as moral compared to when the character is portrayed as immoral (Zhou and Shapiro, 2022).

Furthermore, when individuals experience a sense of merged identity with a character, they may interpret the character's actions as if they were their own and incorporate attributes inferred from those actions into their self-concept (Goldstein & Cialdini, 2007). Such vicarious incorporation has been shown to be stronger when the character is perceived as similar or psychologically close to the individual (Goldstein & Cialdini, 2007). Moreover, the abilities and values attributed to a narrative character can be temporarily integrated into one's own self-concept, which in turn may enhance the individual's SoA (Slater et al., 2014).

Taken together, these findings suggest that similar processes may also occur within VR-based Proteus effect. When an avatar's character is perceived as moral, users tend to feel greater liking and thus identify more readily with that character. Through this identification, the avatar's narrativity and internal attributes may be incorporated into users' self-concepts, leading them to interpret the character's actions as if they were their own.

Additionally, individual differences in specific cognitive abilities may modulate the degree to which the FBI is experienced. One such factor is phenomenological control, defined as the ability to unconsciously modulate one's perceptual experiences in accordance with expectations generated by imaginative suggestions (Lush et al., 2021). Imaginative suggestion refers to the process of inducing experiences such as perceiving a weight that does not actually exist, triggered by suggestions like "a heavy object is resting on your hand." This phenomenon encompasses diverse forms, including paralysis, amnesia, and sensory hallucinations. Previous research (Lush *et al.*, 2020) suggests that the rubber hand illusion (RHI), where one perceives a rubber hand as their own, arises at least partially through phenomenological control. Unlike the RHI, which focuses on a body part, the FBI involves the entire body, yet it may similarly arise through phenomenological control.

Drawing on these considerations, the present study had two primary aims. First, it investigated how an avatar's internal characteristics influence users' SoA, and whether positive narrativity associated with these characteristics enhances SoA within the FBI. These internal characteristics were conveyed through a narrative manipulation, using an avatar with low public familiarity to minimize the influence of pre-existing stereotypes. Second, the study examined whether individual differences in phenomenological control are associated with susceptibility to the FBI.

## 2 Methods

### 2.1 Participants

Thirty-two healthy adults (21.9 ± 1.4 years, mean ± SD) participated in the experiment. This study was approved by the Ethics Committee of the School of Science and Technology at Meiji University. All participants were provided with an explanation of the experimental procedures and safety considerations, and written informed consent was obtained prior to participation.

Individual differences in phenomenological control were assessed using the Japanese version of the Phenomenological Control Scale (PCS-J; Imaizumi & Suzuki, 2025). The PCS-J consists of 12 imaginative suggestions presented auditorily via a personal computer. For each suggestion (e.g., "the hand you are holding out becomes heavy and starts to lower"), participants rated the extent to which they experienced the suggested effect on a six-point Likert scale using a keyboard. Higher scores indicate a greater tendency to experience suggested sensations or actions. The PCS-J assessment was conducted several days prior to the VR experiment in order to use the scores for group assignment.

Participants were assigned to one of two narrative conditions based on their PCS-J scores. The Positive group ($n$ = 16) listened to a positive narrative about the Golem avatar, whereas the Negative group ($n$ = 16) listened to a negative narrative. Following a previous procedure (Lush et al., 2021), group assignment was conducted such that the means and variances of PCS-J scores were approximately matched between the two groups (Positive group: $M$ = 1.81, $SD$ = 0.79; Negative group: $M$ = 1.86, $SD$ = 0.71).

### 2.2 Narrative Manipulation

In this experiment, two narrative versions were created about a Golem avatar, an artificial lifeform with low public familiarity, to induce either positive or negative impressions. Prior to listening to the narratives, participants in both the Positive and Negative groups were provided with a brief, neutral introduction to the Golem. Each narrative described how the Golem responded to a bandit attack on a village. Both narratives were constructed around the same set of psychological traits, such as obedience and emotional composure, as well as identical physical capabilities. In the Positive narrative, these traits and abilities were portrayed as being used to protect villagers from the bandits' assault. In contrast, in the Negative narrative, the same traits and abilities were framed as being misused, resulting in the Golem turning against the villagers and attacking them. Thus, the two narratives differed only in their contextual framing, such that identical internal characteristics led to opposite moral outcomes. The full text of the narratives is provided in the Supplementary Materials.

### 2.3 Apparatus and Experimental Procedure

To present the VR environment and track participants' movements, a head-mounted display and motion-tracking devices were used. Participants wore a head-mounted display (VIVE XR Elite, Q8R100, HTC; resolution: 1920 × 1920 pixels per eye, field of view: 110°, refresh rate: 90 Hz). Participants' movements were tracked using VIVE Trackers (Ultimate, HTC; six degrees of freedom inside-out tracking), which were attached to both wrists. The virtual environment was developed using the Unity game engine (version 2023.2.20f1).

A few days after the PCS-J assessment, participants took part in the main VR experiment. At the beginning of the session, they listened to either a positive or a negative audio narrative about the Golem avatar, depending on their assigned group. The Positive group listened to a positive narrative, whereas the Negative group listened to a negative narrative. The narratives were presented through earphones from a personal computer, and the auditory stimuli were generated using VOICEVOX (version 0.21.1).

After listening to the narrative, participants performed a VR task designed to induce the FBI. Participants experienced the Golem avatar from a first-person perspective. Previous studies have shown that the FBI is enhanced when users observe an avatar whose movements are synchronously coupled with their own movements from a first-person viewpoint (Ohtsuka & Shimada, 2021). A virtual mirror was placed in front of the avatar so that participants could observe the avatar's full body (Figure 1). The VR task lasted for 3 minutes. During the task, participants were instructed to touch and remove red balls that appeared randomly in the virtual space. When the avatar touched a ball, a brief impact sound was presented as auditory feedback.

After completing the VR task, participants completed a series of questionnaires assessing impressions of the avatar, and bodily self-consciousness, and identification with the avatar. The details of each questionnaire and the statistical procedures are described below.

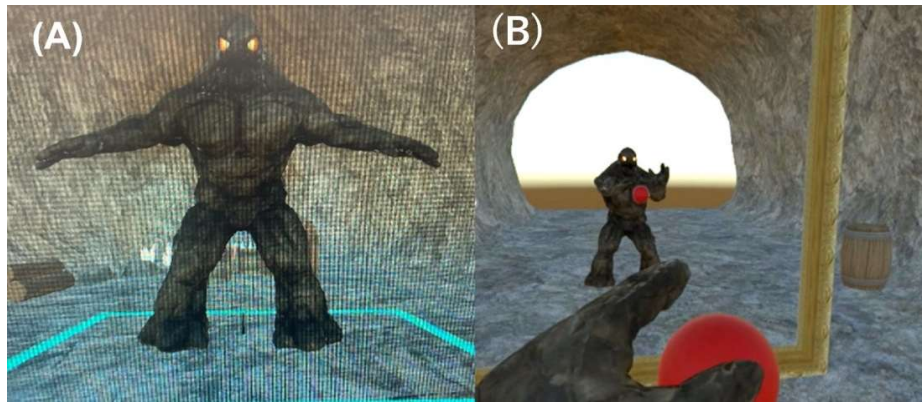

**Figure 1. The Golem avatar and the virtual reality (VR) task used in the experiment.**
(A) The Golem avatar used as the virtual body.
(B) Example of the participant's first-person perspective while controlling the Golem avatar during the task.

### 2.4 Questionnaires and Statistical Analysis

Impressions of the avatar formed through the narrative were assessed using the 20-item Trait Adjective Scale (Hayashi, 1982). Factor scores for personal familiarity, social desirability, and activeness were calculated based on the factor loading matrix reported in a previous study (Hayashi, 1979). Specifically, the item order in the present dataset was aligned with that of the original factor loading matrix, and factor scores were computed as weighted sums using the reported factor loadings.

The degree of the FBI toward the avatar was assessed using a bodily self-consciousness questionnaire adapted from previous studies (Koike & Shimada, 2022). The questionnaire consisted of four items assessing SoO and four items assessing SoA, with each set including two dummy items. All items were rated on a 7-point Likert scale ranging from −3 (strongly disagree) to +3 (strongly agree). For the analyses, only the two non-dummy items for each construct were used, and the mean of these items was calculated as the SoO and SoA scores, respectively.

Psychological closeness and the sense of merged identity with the avatar were assessed using the Inclusion of Other in the Self (IOS) Scale (Aron, Aron, & Smollan, 1992). Participants selected one of eight diagrams depicting two circles with varying degrees of overlap, and the selected level, ranging from 1 to 8, was used as the IOS score, with higher values indicating greater perceived psychological overlap with the avatar.

For all questionnaire measures, scores were computed for each participant and used in the statistical analyses. The normality of each group's score distributions was assessed using the Shapiro–Wilk test. If both groups met the assumption of normality, independent-samples t-tests were conducted; if normality was violated in either group, Mann–Whitney U tests were applied. Within-group associations among questionnaire scores were evaluated using Spearman's rank correlation coefficients. All statistical analyses were conducted using R (version 4.4.2) and Python (version 3.11.7).

## 3 Results

### 3.1 Trait Adjective Scale

Group comparisons on the Trait Adjective Scale revealed significantly higher scores for the Positive group than for the Negative group on all three factors. Specifically, the Positive group scored higher on personal familiarity ($t(25.32) = 5.98$, $p < .001$, $d = 2.12$), social desirability ($t(24.37) = 4.98$, $p < .001$, $d = 1.76$), and activeness ($t(29.11) = 6.17$, $p < .001$, $d = 2.18$) (Figure 2). These results indicate that the narrative manipulation successfully induced more positive impressions of the avatar in the Positive condition.

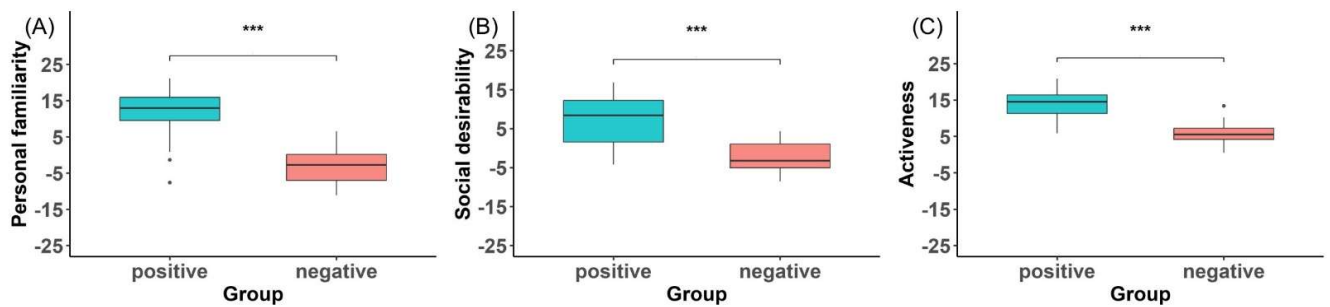

**Figure 2. Group comparisons on the Trait Adjective Scale.**
(A) personal familiarity, (B) social desirability, and (C) activeness scores for the Positive and Negative groups.
The Positive group showed significantly higher scores than the Negative group on all three factors.

## 3.2 Bodily Self-consciousness

For the SoO, one-sample t-tests showed that scores were significantly higher than zero in both the Positive group ($t(15) = 2.53$, $p = .02$, $d = 0.63$) and the Negative group ($t(15) = 2.24$, $p = .04$, $d = 0.56$), indicating that the FBI was successfully induced in both conditions. No significant difference was observed between the groups for SoO scores ($t(29.81) = 0.07$, $p = .95$, $d = 0.02$) (Figure 3A).

For the SoA, scores were significantly higher than zero in both groups. The Positive group showed a significant deviation from zero as indicated by a Wilcoxon signed-rank test ($V = 120$, $p < .001$, $r = 0.88$), and the Negative group also showed significantly higher scores than zero based on a one-sample t-test ($t(15) = 6.45$, $p < .001$, $d = 1.61$). Importantly, the Positive group scored significantly higher than the Negative group on SoA, as revealed by a Mann–Whitney U test ($U = 54.5$, $p = .02$, $r = 0.21$) (Figure 3B).

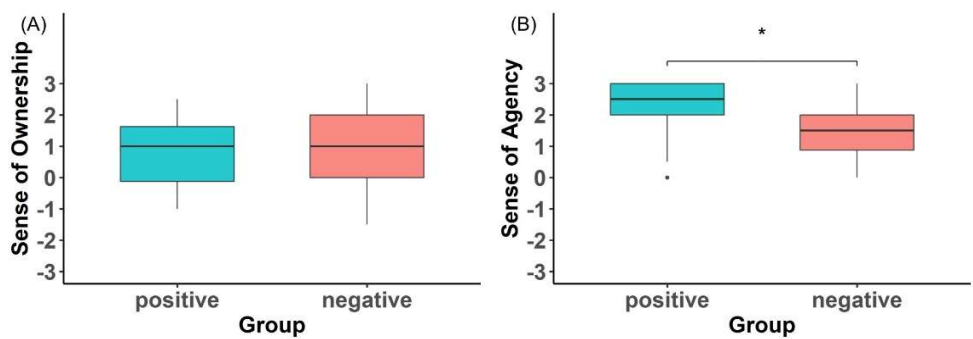

**Figure 3. Group comparisons on sense of ownership (SoO) and sense of agency (SoA).**
(A) SoO scores did not differ between the Positive and Negative groups.
(B) SoA scores were higher in the Positive group than in the Negative group.

## 3.3 IOS scale

The IOS score averaged 4.81 ($SD = 1.68$) in the Positive group and 5.25 ($SD = 2.02$) in the Negative group, indicating moderate levels of perceived psychological overlap with the avatar on the 1–8 scale in both groups. No significant difference was found between the Positive and Negative groups on the IOS Scale, as indicated by a Mann–Whitney U test ($U = −28.5$, $p = .44$, $r = −0.11$). This result suggests that the narrative manipulation did not significantly affect self–avatar identification as measured by the IOS.

## 3.4 Correlation Analyses

Spearman's rank correlation analysis revealed a significant positive association between personal familiarity and SoA scores in the Positive group ($r = 0.51$, $p = .04$), whereas no such association was observed in the Negative group ($r = 0.06$, $p = .81$) (Figure 4). This result indicates that, under the positive narrative condition, greater perceived personal familiarity with the avatar was associated with a stronger SoA.

In contrast, no significant correlations were found between phenomenological control, as measured by PCS-J scores, and bodily self-consciousness measures. Specifically, PCS-J scores were not significantly correlated with SoO scores in either the Positive group ($r = 0.22$, $p = .42$) or the Negative group ($r = 0.06$, $p = .82$), nor with SoA scores in the Positive group ($r = 0.23$, $p = .38$) or the Negative group ($r = 0.17$, $p = .54$).

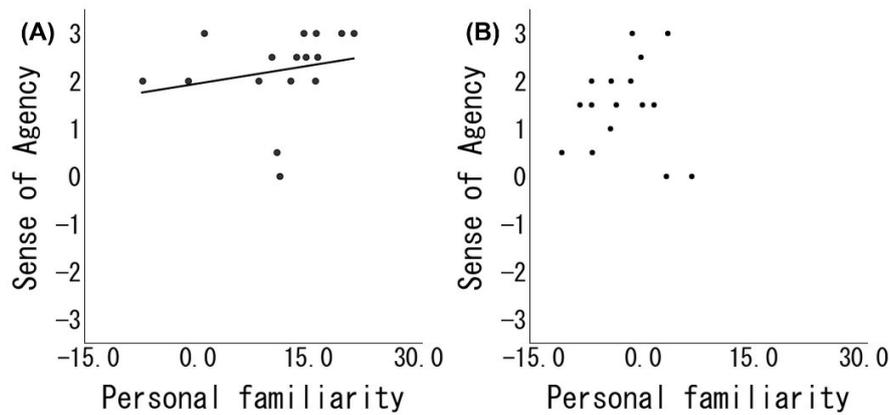

**Figure 4. Correlation between personal familiarity and sense of agency (SoA).**
Scatterplots of personal familiarity and SoA scores for (A) the Positive group and (B) the Negative group.
A positive association was observed in the Positive group, whereas no significant association was observed in the Negative group.

## 4 Discussion

In this study, two research aims were addressed. The first aim was to investigate how positive narrativity attributed to an avatar influences the SoA. The second aim was to examine whether individual differences in phenomenological control are associated with susceptibility to the FBI.

Regarding the first aim, the results of the Trait Adjective Scale demonstrated that the Positive group scored significantly higher than the Negative group on all three factors, namely personal familiarity, social desirability, and activeness. These findings indicate that the narrative manipulation successfully altered participants' impressions of the avatar. Previous studies have shown that moral evaluations of narrative characters strongly influence liking toward those characters (Raney and Bryant, 2002). In the present experiment, participants likely evaluated the morality of the Golem based on the storyline, and the positive narrative, by portraying the avatar as morally favorable, elicited more positive impressions.

These impression changes provide an important basis for understanding the observed modulation of SoA. Prior research suggests that when individuals develop liking for a character through narrative engagement, they tend to internalize the character's attributes and experience stronger identification (Goldstein and Cialdini, 2007). In the present study, participants who perceived the Golem more positively likely experienced enhanced identification, which may have led them to interpret the avatar's actions during the VR task as if they were their own. Consistent with this interpretation, within the Positive group, a significant positive correlation was observed between personal familiarity and SoA. This finding suggests that greater psychological closeness to the avatar was associated with a stronger SoA.

In addition to identification, positive narrativity may have enhanced SoA by shaping participants' expectations about action outcomes. Individuals generally tend to attribute positive outcomes to themselves more readily than negative ones, a tendency often described as a self-serving attribution bias (Mezulis et al., 2004). In motor control tasks, positive outcomes are more likely to be attributed to one's own actions and are associated with stronger SoA than negative outcomes (Oishi, Tanaka, and Watanabe, 2018). Furthermore, positive expectations regarding action outcomes can enhance SoA even in the absence of objectively positive feedback (Gentsch and Synofzik, 2014).

In the present study, the positive narrative portrayed the Golem as producing beneficial outcomes, which may have induced positive expectations regarding the avatar's actions during the virtual reality task. Within the Positive group, a significant positive correlation was observed between personal familiarity and SoA. Previous research has shown that vicarious self-perception is strengthened when a narrative character is perceived as similar or psychologically close to the self (Goldstein and Cialdini, 2007). Consistent with this view, participants who felt greater personal familiarity with the avatar may have been more inclined to attribute its actions to themselves. This psychological closeness, together with positive expectations shaped by the narrative, could have facilitated the attribution of the avatar's movements to oneself, thereby enhancing SoA. Taken together, these findings suggest that narrativity functions as a contextual factor operating at a higher cognitive level, shaping agency attribution during avatar embodiment in addition to sensorimotor cues such as visuomotor synchrony.

Despite the observed effects on SoA, no significant difference was found between the Positive and Negative groups on the Inclusion of Other in the Self scale. Although this scale was intended to assess psychological closeness and merged identity with the avatar, participants may have interpreted it primarily in terms of bodily overlap rather than psychological identification within the context of a full body embodiment task. Given that participants directly experienced the avatar's body as their own during the VR task, judgments on the scale may have been biased toward physical alignment rather than narrative based identification.

Turning to the second aim of the study, no significant associations were found between phenomenological control and either SoO or SoA within the FBI. This result contrasts with previous findings showing that phenomenological control is positively associated with SoO in the RHI (Lush et al., 2020). In the rubber hand illusion, the experience relies heavily on the integration of visual and tactile cues applied to a single body part. Phenomenological control may play a substantial role in shaping this form of ownership.

In contrast, the FBI in the present study was induced under conditions of strong visuomotor synchrony, with participants observing the avatar's movements as their own via a virtual mirror. Previous work has shown that mirror-based visuomotor coupling robustly enhances SoO (Gonzalez-Franco et al., 2010). Under such conditions, the FBI may be more readily induced, thereby reducing the contribution of individual differences in phenomenological control. This difference in task structure may explain why phenomenological control was not significantly related to bodily self-consciousness in the present study.

The findings of this study have important implications for understanding how virtual reality environments can modulate agency and behavior. The observation that positive narrativity enhances

SoA suggests that narrative context can be strategically used to promote a stronger sense of control over one's actions in virtual environments. Enhanced SoA has been linked to improved action regulation (Kaiser et al., 2021) and to better motor performance in illusion based paradigms such as the rubber hand illusion (Matsumiya, 2021). Accordingly, positive narrative framing may serve as an effective means of facilitating agency and supporting adaptive behavior in VR applications.

Importantly, the enhancement of SoA observed in this study did not depend on individual differences in phenomenological control. This suggests that narrativity based modulation of agency may operate relatively independently of such individual traits, at least under conditions of strong visuomotor coupling. Overall, the present findings highlight narrativity as a powerful contextual factor that shapes embodiment and agency in VR, offering new insights into how virtual experiences can be designed to influence cognition and behavior.

**References**

Aron, A., Aron, E.N., Smollan, D. (1992). Inclusion of other in the self scale and the structure of interpersonal closeness. *J. Pers. Soc. Psychol*. 63, 596–612. doi:10.1037/0022-3514.63.4.596

Cohen, J. (2018). "Defining identification: A theoretical look at the identification of audiences with media characters". in *Advances in Foundational Mass Communication Theories,* ed. R. Wei (London: Routledge), 253–272. doi:10.4324/9781315164441-14

Gallagher, S. (2000). Philosophical conceptions of the self: Implications for cognitive science. *Trends. Cogn. Sci.* 4, 14–21. doi:10.1016/s1364-6613(99)01417-5

Gentsch, A., Synofzik, M. (2014). Affective coding: The emotional dimension of agency. *Front. Hum. Neurosci.* 8:608. doi:10.3389/fnhum.2014.00608

Goldstein, N.J., Cialdini, R.B. (2007). The spyglass self: A model of vicarious self-perception. *J. Pers. Soc. Psychol.* 92, 402–417. doi:10.1037/0022-3514.92.3.402

Gonzalez-Franco, M., Pérez-Marcos, D., Spanlang, B., Slater, M. (2010). "The contribution of real-time mirror reflections of motor actions on virtual body ownership in an immersive virtual environment". in *2010 IEEE Virtual Reality (VR),* 111–114. doi:10.1109/vr.2010.5444805

Hayashi, F. (1979). A multidimensional analytical to the measurement of individual differences in interpersonal cognitive structure (4): an application of the INDSCAL model. *Jpn. J. Psychol.* 50, 211–218.

Hayashi, F. (1982). The measurement of individual differences in interpersonal cognitive structure (8): relationships between cognitive dimensions and some personality variables. *Jpn. J. Psychol.* 22, 1–9.

Imaizumi, S., Suzuki, K. (2025). The Japanese version of the Phenomenological Control Scale. *Neurosci. Conscious.* 2025. doi:10.1093/nc/niaf008

Kaiser, J., Buciuman, M., Gigl, S., Gentsch, A., Schütz-Bosbach, S. (2021). The interplay between affective processing and sense of agency during action regulation: A review. *Front. Psychol.* 12:716220. doi:10.3389/fpsyg.2021.716220

Kilteni, K., Groten, R., Slater, M. (2012). The sense of embodiment in virtual reality. *Presence Teleoperators Virtual Environ.* 21, 373–387. doi:10.1162/pres_a_00124

Koike, Y., Shimada, S. (2022). The effects of the full-body illusion to an intelligent avatar on executive function and personality trait. *J. Virtual Reality Soc. Japan.* 27, 385–392.

Lenggenhager, B., Tadi, T., Metzinger, T., Blanke, O. (2007). Video ergo sum: Manipulating bodily self-consciousness. *Science* 317, 1096–1099. doi:10.1126/science.1143439

Lush, P., Botan, V., Scott, R. B., Seth, A. K., Ward, J., Dienes, Z. (2020). Trait phenomenological control predicts experience of mirror synaesthesia and the Rubber Hand Illusion. *Nat. Commun.* 11:4853. doi:10.1038/s41467-020-18591-6

Lush, P., Scott, R. B., Seth, A. K., Dienes, Z. (2021). The Phenomenological Control Scale: Measuring the capacity for creating illusory nonvolition, hallucination and delusion. *Collabra. Psychol.* 7:29542. doi:10.1525/collabra.29542

Matsumiya, K. (2021). Awareness of voluntary action, rather than body ownership, improves motor control. *Sci. Rep.* 11:75. doi:10.1038/s41598-020-79910-x

Mezulis, A. H., Abramson, L. Y., Hyde, J. S., Hankin, B. L. (2004). Is there a universal positivity bias in attributions? A meta-analytic review of individual, developmental, and cultural differences in the self-serving attributional bias. *Psychol. Bull.* 130, 711–747. doi:10.1037/0033-2909.130.5.711

Oishi, H., Tanaka, K., Watanabe, K. (2018). Feedback of action outcome retrospectively influences sense of agency in a continuous action task. *PLoS One* 13:e0202690. doi:10.1371/journal.pone.0202690

Otsuka, I., Shimada, S. (2021). The effect of delayed visual feedback on the full-body illusion using VR. *J. Jpn. Soc. Fuzzy Theory Intell. Inform.* 33, 657–662.

Raney, A.A. (2002). Moral judgment and crime drama: An integrated theory of enjoyment. *J. Commun.* 52, 402–415. doi:10.1093/joc/52.2.402

Slater, M.D., Johnson, B. K., Cohen, J., Comello, M. L. G., Ewoldsen, D. R. (2014). Temporarily expanding the boundaries of the self: Motivations for entering the story world and implications for narrative effects. *J. Commun.* 64, 439–455. doi:10.1111/jcom.12100

van Krieken, K., Hoeken, H., Sanders, J. (2017). Evoking and measuring identification with narrative characters – a linguistic cues framework. *Front. Psychol.* 8:1190. doi:10.3389/fpsyg.2017.01190

Yee, N., Bailenson, J. (2007). The Proteus effect: The effect of transformed self-representation on behavior. *Hum. Commun. Res.* 33, 271–290. doi:10.1111/j.1468-2958.2007.00299.x

Yoon, G., Vargas, P.T. (2014). Know thy avatar. *Psychol. Sci.* 25, 1043–1045. doi:10.1177/0956797613519271

Zhou, S., Shapiro, M.A. (2022). Impacts of character morality on egocentric projection and identification. *Poetics* 95:101731. doi:10.1016/j.poetic.2022.101731